\documentclass[aps,prl,a4,showpacs,tighten]{revtex4}
\begin{document}

\title{Post Newtonian Rigid Body}

\author{Chongming Xu$^1$}
\email{cmxu@njnu.edu.cn}
\author{Jin-he Tao$^{2,3}$}
\author{Xuejun Wu$^1$}
\affiliation{$^1$Department of Physics,
Nanjing Normal University, Nanjing 210097, P.R. China \\
$^2$Purple Mountain Observatory, The Chinese Academy of Sciences,
Nanjing 210008, P.R. China \\
$^3$National Astronomical Observatories, The Chinese Academy of Sciences, P.R. China}

\date{\today}

\begin{abstract}
\noindent In this paper, it is the first time to construct a
complete post-Newtonian (PN) model of a rigid body by means of a
new constraint on the mass current density and mass density. In
our PN rigid body model most of relations, such as spin vector
proportional to the angular velocity, the definition on the moment
of inertia tensor, the key relation between the mass quadrupole
moment and the moment of inertia tensor, rigid rotating formulae
of mass quadrupole moment and the moment of inertia tensor, are
just the extension of the main relations in Newtonian rigid body
model. When all of $1/c^2$ terms are neglected, the PN rigid body
model and the corresponding formulae reduce to Newtonian version.
The key relation is obtained in this paper for the first time,
which might be very useful in the future application to problems
in geodynamics and astronomy.
\end{abstract}

\maketitle

The idea of Newtonian rigid body has been used to treat the
rotation of astronomical bodies up to now. For example, most
Newtonian treatments of Earth's rotation are based on an accurate
rigid body theory (such as SMART 97 \cite{bret97,bret98}) plus the
perturbative argument from elasticity, the oceans, the atmosphere,
the core and so on \cite{deha99}. The idea of a rigid body in
Newtonian theory is very powerful in introducing three principal
axes of a body, the spin axis, rotation axis and figure axis
without ambiguity. Then it makes the problem much simpler since
there exists the key simple relation between the quadrupole moment
and the moment of inertia tensor in a rigid body. But even in
Newtonian theory the concept of a rigid body is only an ideal one,
because there is no real rigid body in the physical world
\cite{gold80}. Owing to the modern high accuracy requirement, the
Newtonian theory has to be replaced by Einstein's general
relativity (at least its first post-Newtonian (PN) approximation).
The problems of the post-Newtonian rigid body have been discussed
for a long time since Born's kinematical rigidity (see Dixon's
review \cite{dixo79}). The kinematical rigidity is dependent on
the internal velocity distribution within the body while not
considering the stress and energy fluxes contribution on the
energy-momentum tensor $T^{\alpha\beta}$. Dixon \cite{dixo79},
Thorne \& G\"ursel \cite{thor83}, Klioner
\cite{klio96,klio97,klio01} and Soffel \cite{soff94} have done in
a much better way, so-called dynamical rigidity way, in which
$T^{\alpha\beta}$ of the body and the gravitational field caused
by the body satisfy some certain interdependency. The
interdependency is not the same for different authors. However all
of the different interdependencies make the PN spin $S^i$
proportional to the angular velocity $\Omega^a$ and define a
relativistic moment of inertia tensor. Certainly, the
interdependency for the Newtonian rigid body just corresponds to
the simple formula between mass density and current density (shown
in Eq.~(\ref{3gt})). But none of them obtains the key simple
relation between the PN quadrupole moment $M_{ab}$ and the PN
moment of inertia tensor $I_{ab}$ like Newtonian one as shown in
Eq.~(\ref{6gt}). Some scientists even assert that such a key
relation is invalid in general relativity \cite{thor83,klio96}.
Therefore the idea of PN rigidity almost has not been directly
applied to practical problems up to now. We have a different
opinion. We think because no one has discovered a suitable
interdependency inside the energy-momentum tensor and the
gravitational field before, then the key relation between PN
$M_{ab}$ and PN $I_{ab}$ is not found. In this paper we present a
suitable new interdependency to obtain the PN $I_{ab}$ and the key
relation between PN $M_{ab}$ and PN $I_{ab}$ similar to Newtonian
one. It is the first time to solve the rigid body problem on the
post-Newtonian level. Recently we suggested another
interdependency inside the energy-momentum tensor and
gravitational field of the rigid body on PN level by means of a
special gauge condition \cite{tao03}, but the special gauge
condition is more or less speculative and difficult to be commonly
accepted. In this paper we totally discard the special gauge
condition, and present a general expression of the rigid spin with
or without external field (free precession).

First let us recall the basic aspects of the Newtonian rigid body.
We take $\Sigma$ and $\Sigma^a$ $(=\Sigma V^a)$ as the mass
density and the mass current density of a rigid body $A$
respectively. Then the mass multipole moments $M_L$ and spin $S^a$
of body $A$ are defined as $ M_L=\int_Ad^3X \Sigma \hat{X}^L$ and
$ S^a=\epsilon_{abc}\int_Ad^3XX^b \Sigma^c$, where $\hat{X}^L$ is
an abbreviation of $X^{<i_1}X^{i_2} \cdots X^{i_l>}$, in which
$i_j$ ($j=1, 2, \cdots, l$) is a spatial index, the angular
brackets mean ``symmetrize and take the trace-free part'' (STF).
In Newtonian rigid body the rotational angular velocity $\bf
\Omega$ is independent of spatial coordinate, we have
\begin{equation}\label{3gt}
\Sigma^a=\epsilon_{abc}\Sigma \Omega^bX^c \, .
\end{equation}
Insert Eq.~(\ref{3gt}) into the expression of spin, we have the
linear relation between the spin and angular velocity $
S^a=I_{ab}\Omega^b $, where the moment of inertia tensor $I_{ab}$
is $ I_{ab}=I_{ba}=\int_Ad^3X \Sigma(X^2 \delta_{ab}-X^aX^b)$.

The mass quadrupole moment and the moment of inertia tensor
satisfy the key relation
\begin{equation}\label{6gt}
  M_{ab}=-I_{ab}+\frac{1}{3}\delta_{ab}I_{cc} \, .
\end{equation}
By means of the continuity equation $ \partial_T \Sigma
+\partial_a \Sigma ^a=0 $, the time derivative of the moment of
inertia tensor $I_{ab}$ is proportional to the angular velocity
$\bf \Omega$
\begin{equation}\label{9gt}
  \dot{I}_{ab}\equiv \frac{d}{dT}I_{ab}
  =(\epsilon _{apq}I_{qb}+\epsilon _{bpq}I_{aq})\Omega^p \, .
\end{equation}
$\dot{M}_{ab}$ satisfies a similar relation as (\ref{9gt}).
Therefore, $I_{ab}$ and $M_{ab}$ like constant tensors, will
rigidly rotate in space with the angular velocity $\bf \Omega$.

When we discuss PN rigidity, we will use the symbols, notations
and conventions following the 1PN theoretical framework presented
by Damour, Soffel and Xu (cited below as the DSX scheme
\cite{damo91,damo92,damo93}) since the DSX scheme is not only
rather simple and complete but also describes the 1PN definition
of spin in a satisfactory manner. In the DSX scheme a complete 1PN
general relativistic celestial mechanics for $N$ arbitrarily
composed and shaped, rotating deformable bodies is described. Here
we will briefly summarize the notation and definition in the DSX
scheme. In the post Newtonian expansion we will always abbreviate
the order symbol $O(c^{-n})$ simply as $O(n)$. A spatial
multi-index containing $l$ indices is simply denoted by $L$ (and
$K$ for $k$ indices, etc), i.e. $L \equiv i_1 i_2 \cdots i_l$. A
multisummation is always understood for repeated multi-indices
$S_L T_L = \sum_{i_1}\sum_{i_2} \cdots \sum_{i_l} S_{i_1i_2 \cdots
 i_l} T_{i_1 i_2 \cdots i_l}$. Given a spatial vector, $n^i$, its
$l$th tensorial power is denoted by $n^L \equiv
n^{i_1}n^{i_2}\cdots n^{i_l}$. Also, $\partial _L \equiv
\partial_{i_1}\partial_{i_2}\cdots\partial_{i_l}$. Besides angular
brackets the symmetric and trace-free (STF) part of a spatial
tensor will be denoted by a caret when no ambiguity arises: ${\rm
STF} _{i_1\cdots i_l} (T_{i_1} \cdots T_{i_l}) \equiv  T_{<i_1
\cdots i_l>} = \hat{T}_L$. The spatial indices, $i,j=1,2,3$ are
freely raised or lowered by means of the Cartesian metric $\delta
_{ij}= \delta^{ij}= diag (+1,+1,+1)$ in Cartesian coordinates. The
metric is presented by means of potential $W$ and vector potential
$W_a$ (see Eq.~(4.1) of Ref. \cite{damo91}). $W$ and $W_a$ can be
separated into self-part (with a ``+'') and external-part (with a
bar), i.e. $W=W^+ + \overline{W}$ and $W_a=W_a^+ +
\overline{W}_a$. The self-part $W^+$ and $W^+_a$ will be solved by
the gravitational mass density $\Sigma$ and the mass current
density $\Sigma^a$ as: $\Sigma \equiv (T^{00}+T^{ss})/c^2$, and
$\Sigma ^a \equiv T^{0a}/c $, through the 1PN Einstein field
equation and the coordinate conditions (gauge conditions) (see
Eq.~(4.3) of Ref. \cite{damo91}), where $T^{\alpha\beta}$ is the
energy-momentum tensor. $W^+$ and $W^+_a$ will be expanded by STF
BD mass moments $M_L$ and STF spin moments $S_L$ (see Eq.~(6.11)
of Ref. \cite{damo91}). The external part $\overline{W}$ and
$\overline{W}^a$ can be expanded by gravito-electric tidal moments
$G_L$ and gravito-magnetic tidal moments $H_L$ (see Eq.~(4.15) of
Ref. \cite{damo92}). $G_L$ and $H_L$ are also STF spatial tensors
dependent on time only. BD mass moments \cite{blan89} are widely
accepted as the best PN mass moments, which have the form
\begin{eqnarray}\label{11gt}
M^A_L (T) & \equiv & \int_A d^3X\hat{X}^L \Sigma
  + \frac{1}{2(2l+3)c^2} \frac{d^2}{dT^2}\left[
  \int_A d^3X \hat{X}^L X^2 \Sigma \right] \nonumber \\
  && - \frac{4(2l+1)}{(l+1)(2l+3)c^2} \frac{d}{dT}
  \left[ \int_A d^3X \hat{X}^{aL} \Sigma ^a \right]
  \quad (l \geq 0) \, .
\end{eqnarray}
The PN spin moment has been discussed for a long
time \cite{damo93,damo91a}. In the DSX scheme, the expression of
1PN spin of body $A$ is
\begin{eqnarray}\label{12gt}
S^{A,~{\rm PN}}_a &  \equiv &  \epsilon_{abc}
  \int_Ad^3X X^b \left[\Sigma^c(1+\frac{4}{c^2}W^A)-\frac{4}{c^2}
  \Sigma( W_c^{+A} +\frac{1}{8}\partial_c\partial_T Z_A^+) \right] \nonumber \\
&&  + \frac{1}{c^2}\sum_{l\geq 0} \frac{1}{l!}
  \left[\frac{1}{2l+3}H^A_{aL}\hat{N}^A_L -\frac{l}{l+1}M^A_{aL}H^A_L
  \right] \nonumber \\
&&  -\ \epsilon_{abc}\frac{1}{2c^2}
  \sum_{l\geq 0}\frac{1}{l!(l+2)(2l+5)}\left[ (l+10)\hat{N}^A_{bL}{\dot
  G}^A_{cL}\right. \nonumber \\
&&    \left. +\ 8(2l+3)\hat{P}^A_{bL}G^A_{cL}-
  (l+2){\dot N}^A_{<bL>}G^A_{cL}\right] + O(4) \, ,
\end{eqnarray}
where $Z^+ _A \equiv G \int_Ad^3X' \Sigma(T_A,~ {\bf X}')
 |{\bf X}-{\bf X'}|$, $W^{+A}_c \equiv G \int_Ad^3X' \Sigma^c(T_A,~{\bf X}')/
 |{\bf X}-{\bf X'}|$, dot means time derivative
 $\partial _T$, $\hat{N}_L$ and $\hat{P}_L $ are defined as
 (see Eq.~(2.10) of \cite{damo93})
$\hat{N}_L \equiv \int_A d^3X X^2\hat{X}^L
  \Sigma $ and $\hat{P}_L \equiv \int_A d^3X \hat{X}^{aL}
  \Sigma ^a $ respectively.
Later we omit to indicate the body label, $A$, on all quantities.
In Ref. \cite{damo93}, $\hat{N}_L$ and $\hat{P}_L$ are called
``bad moments''. In such a definition of 1PN spin vector, $S^{\rm
PN}_a$ satisfies the rotational equation of motion (see Eq.~(3.11)
of Ref. \cite{damo93}). We also have 1PN continuity equation for
($\Sigma ,~\Sigma^a$) (see Eq.~(5.6b) of Ref. \cite{damo91})
\begin{equation} \label{16gt}
\partial_T \Sigma+\partial_a \Sigma^a=\frac{1}{c^2}
(\partial_TT^{bb}-\Sigma \partial_TW) + O(4) \, .
\end{equation}
Those are the equations which are taken from the DSX scheme and
will be used in the following discussion on the PN rigidity.

The definition of the 1PN rigid body has to agree with the
Newtonian rigid body when $1/c^2$ terms are neglected. In the 1PN
rigid body the angular velocity should be independent of the local
coordinate $X^a$ of body $A$. In the DSX scheme we substitute
$\Sigma$ and $\Sigma^c$ for the energy-momentum tensor
$T^{\alpha\beta}$, therefore the interdependency inside the
energy-momentum tensor and gravitational field in Ref.
\cite{thor83,klio96,klio97} might be replaced by inside
$\Sigma,~\Sigma^c$ and gravitational field. We expect that the
interdependency will produce equations similar to (\ref{6gt}) and
(\ref{9gt}) on the PN level. On the PN level it is sufficient to
replace $T^{\alpha\beta}$ by $\Sigma ,~\Sigma^c$ and their
derivative \cite{xu01}. Before a further discussion on the rigid
body, the rigid BD moments and the rigid 1PN spin vector should be
considered. In the rigid body, since $\hat{P}_L$ and $\hat{N}_L$
are PN terms we can substitute the Newtonian relations
(Eq.(\ref{3gt}) and Newtonian continuity equation) for the
definitions of $\hat{N}_L$ and $\hat{P}_L$. It is easy to prove
\begin{equation}\label{17gt}
P_{<L>} = - \frac{1}{2l+1} \dot{N}_{<L>} \, .
\end{equation}

\noindent{\bf Lemma 1} \hspace*{0.5cm} The rigid BD mass moments
of rigid body $A$ (Eq.~(\ref{11gt})) can be simplified as
\begin{equation}\label{18gt}
M_{L} = \int_A d^3X X^{<L>} \left[ \Sigma + \frac{1}{c^2}
  \left( \frac{l+9}{2(l+1)(2l+3)} \right) X^2 \ddot{\Sigma}
  \right] + O(4)\, .
\end{equation}

\noindent Proof of Lemma 1. \hspace*{0.3cm}Starting from
Eq.~(\ref{11gt}) (the definition of BD mass moment) and replacing
the third term in the RHS of Eq.~(\ref{11gt}) by Eq.~(\ref{17gt}).
The prove will be carried out directly. In fact only the
relativistic quadrupole moment is interested in, because in the
solar system all of the {\bf relativistic} higher multipole
moments are too small to be considered in any modern measurement
in the anticipation of the day. Therefore we only take the
relativistic quadrupole moments into account:
\begin{equation}\label{19gt}
M_{ab} = \int_A d^3X X^{<ab>} \left( \Sigma + \frac{11}{42c^2} X^2
\ddot{\Sigma} \right) \, .
\end{equation}
We define the PN part of the mass density $\Sigma_{\rm PN} \equiv
 \frac{11}{42} X^2 \ddot{\Sigma} $ and the total mass density
\begin{equation}\label{20gt}
\overline{\Sigma} \equiv  \Sigma
  + \frac{\Sigma_{\rm PN}}{c^2} \, .
\end{equation}

\noindent {\bf Lemma 2} \hspace*{0.5cm}The rigid PN spin vector of
body $A$ (see Eq.~(\ref{12gt})) can be reduced to
\begin{eqnarray}\label{21gt}
S_a^{\rm PN} &=& \epsilon _{abc} \int_A d^3X X^b \left\{
  \Sigma ^c + \frac{\Sigma}{c^2} \left( \frac{7}{2}
  \epsilon_{cde}\Omega^d \partial _e Z^+
  + {1 \over 2}\epsilon_{edf}\Omega^d X^f\partial _{ce}Z^+
  \right) \right. \nonumber \\
  && \left. +\ \frac{1}{c^2} \sum _{l \geq 0} \frac{\Sigma}{l!}
  \left[ 4\epsilon_{cde} \Omega^d X^eX^{<L>} G_L(T)
  + \frac{1}{l+2}\epsilon_{ced} X^{<dL>} H_{eL} \right. \right.
  \nonumber \\
  && \left. \left. -\ \frac{l+10}{2(l+2)(2l+5)} \hat{X}^L X^2
  \dot{G}_{cL} + \frac{l+10}{2(l+2)(2l+5)}\partial_T(\ln \Sigma)
  \hat{X}^L X^2 G_{cL} \right] \right\} \, .
\end{eqnarray}
Substituting Eqs.~(\ref{3gt}) and (\ref{17gt}) for the PN part of
Eq.~(\ref{12gt}), integrating by parts, using the surface
integration for whole body $A$ to be zero and taking some STF
formulae, we obtained Eq.~(\ref{21gt}).

We define the PN self-part and PN external part of current density
as
\begin{eqnarray}
\Sigma^c_{\rm self} &\equiv& \Sigma \left( \frac{7}{2}
  \epsilon_{cde}\Omega^d \partial _e Z^+
  + {1 \over 2}\epsilon_{edf}\Omega^d X^f\partial _{ce}Z^+
  \right) \, , \label{22gt} \\
\Sigma ^c_{\rm ext} &\equiv& \sum _{l \geq 0} \frac{\Sigma}{l!}
  \left[ 4\epsilon_{cde} \Omega^d X^eX^{<L>} G_L(T)
  + \frac{1}{l+2}\epsilon_{ced} X^{<dL>} H_{eL} \right.
  \nonumber \\
  &&  \left. - \frac{l+10}{2(l+2)(2l+5)} \hat{X}^L X^2
  \dot{G}_{cL} + \frac{l+10}{2(l+2)(2l+5)}\partial_T(\ln \Sigma)
  \hat{X}^L X^2 G_{cL} \right] \, . \label{23gt}
\end{eqnarray}
Both of $\Sigma ^c_{\rm self}$ and $\Sigma^c_{\rm ext}$ as well as
$\Sigma^c$ itself are spatially compact-supported. When tidal
moments $G_L$ and $H_L$ are equal to zero, $\Sigma^c$ and
$\Sigma^c_{\rm self}/c^2$ form the PN self-part of spin vector. We
can define
\begin{equation}\label{24gt}
\overline{\Sigma}^c \equiv \Sigma ^c + \frac{\Sigma^c_{\rm
self}}{c^2}
  + \frac{\Sigma^c_{\rm ext}}{c^2}\, ,
\end{equation}
then Eq.~(\ref{21gt}) becomes 
\begin{equation}\label{25gt}
S^{\rm PN}_a = \epsilon_{abc} \int_A d^3 X X^b
 \overline{\Sigma }^c \, .
\end{equation}
By comparing the Newtonian definition of spin with
Eq.~(\ref{25gt}), $\overline{\Sigma}^c$ is the fully PN quantity.

We add $ \frac{1}{c^2}\left[ \partial_T (\Sigma_{\rm PN}) +
\partial_a \left(\Sigma ^a_{\rm self} + \Sigma ^a_{\rm ext}
  \right) \right] $ to both sides of Eq.~(\ref{16gt}) and have
\begin{equation}\label{26gt}
\partial_T \overline{\Sigma} + \partial_a \overline{\Sigma}^a =
  \frac{1}{c^2}\left[ \partial_T T^{bb} - \Sigma \partial_T W
  + \partial_T \Sigma_{\rm PN} + \partial_a(\Sigma^a_{\rm self}
  + \Sigma^a_{\rm ext}) \right] \, .
\end{equation}

Now we construct the model of the PN rigid body by constraint on
$\overline{\Sigma}^c$ and $\overline{\Sigma}$ to satisfy
\begin{equation} \label{27gt}
\overline{\Sigma}^a + \frac{1}{2c^2}X^a\left[ \partial_TT^{bb}
  -\Sigma \partial_TW+\partial_T \Sigma_{\rm PN}
  +\partial_a( \Sigma ^a_{\rm self} + \Sigma^a_{\rm ext}) \right]
  =\epsilon_{abc}\Omega^bX^c \overline{\Sigma}+O(4)\, .
\end{equation}
The relation (Eq.(\ref{27gt})) is our most important assumption
for PN rigid body. Considering that $\overline{\Sigma}^c$ and
$\overline{\Sigma}$ are expressed by $\Sigma ^c$ and $\Sigma$,
which are related to $T^{\alpha\beta}$ in the DSX scheme, then
Eq.~(\ref{27gt}) is also constraint to $T^{\alpha\beta}$. When
$1/c^2$ terms are neglected, Eq.~(\ref{27gt}) goes to
Eq.~(\ref{3gt}). Later we will see that only in this model PN mass
quadrupole moments and the moment of inertia tensors satisfy the
similar Newtonian key relations as in Eq.~(\ref{6gt}). We were not
surprised with the appearance of the time derivative of $\Sigma $
in Eq.~(\ref{27gt}), since in the DSX scheme $T^{\alpha\beta}$ can
be fully represented by $\Sigma$ and $\Sigma^a$ and their
derivatives without difficulty \cite{xu01}. In the
interdependencies described by Thorne and G\"ursel \cite{thor83}
and Klioner \cite{klio97}, they have their own models of the rigid
body by another constraint to $T^{\alpha\beta}$. By comparing the
constrained equations (to see Eq.(A7) in \cite{thor83} (or Eq.(7)
in \cite{klio97}) and Eq.(8) in \cite{klio97}) with
Eq.~(\ref{27gt}), we see that Eq.~(\ref{27gt}) is more
complicated, but still reasonable.

Substituting Eq.~(\ref{27gt}) for Eq.~(\ref{25gt}), we obtain the
linear relation between the PN spin vector of the rigid body and
the angular velocity:
\begin{equation}\label{30gt}
S^{\rm PN}_a=I_{ab}\Omega^b+O(4)\, ,
\end{equation}
where the moment of inertia tensor is
\begin{equation}\label{31gt}
I_{ab}=I_{ba}=\int_Ad^3X(\delta_{ab}{\bf X}^2-X^aX^b)
 \overline{\Sigma} + O(4) \, ,
\end{equation}
in which $\overline{\Sigma}$ is defined in Eq.~(\ref{20gt}).

\noindent
By comparing Eq.~(\ref{31gt}) with Eq.~(\ref{19gt}), we
have
\begin{equation}\label{32gt}
  M_{ab} = -I_{ab}+\frac{1}{3}\delta_{ab}I_{cc}+O(4)\, .
\end{equation}
Eq.~(\ref{32gt}) is the key relation between the PN mass
quadrupole moment (rigid BD moment) and the PN moment of inertia
tensor. It is just this relation that makes the model of the rigid
body very useful and applicable on the PN level as shown in the
Newtonian case. This is the first time that we have obtained the
PN key relation in this paper. Making use of the extended 1PN
continuity equation Eq.~(\ref{26gt}), we immediately have
\begin{equation} \label{33gt}
\dot{I}_{ab}\equiv \frac{d}{dT}I_{ab}
  =(\epsilon _{apq}I_{qb}+\epsilon _{bpq}I_{aq})\Omega^p
  + O(4) \, .
\end{equation}
PN $\dot{M}_{ab}$ satisfies a similar relation as (\ref{33gt}).
From Eq.~(\ref{33gt}) the behavior of the PN $I_{ab}$ (and also PN
$M_{ab}$) in our rigid model are just like the Newtonian version
(Eq.~(\ref{9gt})), i.e. $I_{ab}$ and $M_{ab}$ rigidly rotate as a
whole.

Eq.~(\ref{33gt}) means that we always can introduce a rotation
matrix $P_{ia}(T)$, which is a time-dependent orthogonal matrix
and transform the PN reference system (RS) to the corotating
reference system with rigid body (RS$^+$). $P_{ia}(T)$ can be
constructed by the rotational angular velocity $\Omega^a$ of rigid
body according to the relation:
$\Omega^a(T)=\frac{1}{2}\epsilon_{abc}$$P_{ib}(T)\dot{P}_{ic}(T)$
(\cite{brum95}). In the new corotating coordinates we get
\begin{equation}\label{33agt}
\frac{d \widetilde{I}_{ij}}{dT} = O(4) \, ,
\end{equation}
where $\widetilde{I}_{ij} = P_{ia}P_{jb}I_{ab}$. $P_{ia}$
satisfies the following relations: $P_{ia}P_{ja} = \delta_{ij}$,
$P_{ia}P_{ib} =\delta_{ab}$ and
$dP_{ia}/dT$$=\epsilon_{abc}$$\Omega^b$$P_{ic}$ (here we use
$\Omega^b$ to substitute for $\omega^j$ in \cite{brum95}). The
proof is easy by means of (\ref{33gt}). Eq.~(\ref{33agt}) shows it
is possible to introduce the PN Tisserand reference system.

At last, we should emphasize that the calculation of the PN moment
of inertia tensor Eq.~(\ref{31gt}) is not too difficult, although
the constraint relation on $\overline{\Sigma}^c$ and
$\overline{\Sigma}$ in Eq.~(\ref{27gt}) in the model of the PN
rigid body is complicated. In practical problems, from our PN
rigid spin, the PN moment of inertia tensor (Eqs.~(\ref{30gt})--
(\ref{32gt})) it is possible to define the three principal axes of
the body, the spin axis, rotation axis and figure axis as
described in the Newtonian theory, which we will discuss in a
separate paper in the future.

In conclusion, the rigid BD (PN) mass multipole moments
Eq.~(\ref{18gt}) and the rigid PN spin moment Eq.~(\ref{25gt}) are
discussed in this paper. We successfully have constructed a new PN
model of the rigid body in which the constraint on
$\overline{\Sigma}^c$ and $\overline{\Sigma}$ satisfies
Eq.~(\ref{27gt}). Our PN rigid body model will reduce to the
Newtonian one when all of $1/c^2$ terms are neglected. Most of
relations in our PN rigid body model, such as the spin vector
proportional to the angular velocity ${\bf \Omega}$
(Eq.~(\ref{30gt})), the definition on the moment of inertia tensor
(Eq.~(\ref{31gt})), the key relation between the mass quadrupole
moment and the moment of inertia tensor (Eq.~(\ref{32gt})), the
rigidly rotating formulae of $I_{ab}$ and $M_{ab}$ (to see
Eq.~(\ref{33gt})) are similar to the Newtonian rigid body model
where the corresponding relations are mentioned at the beginning
of this paper. Especially, the PN key relation between $M_{ab}$
and $I_{ab}$ might be applied to the practical problems in
geodynamics and astronomy in the future, e.g. the discussion on
the relativistic effects of the nutation and precession.

This work was supported by the National Natural Science Foundation
of China (Grant Nos. 10273008 and 10233020). We would like to
thank Prof. T.-Y. Huang for his helpful discussion.



\end{document}